\documentclass[letter,twocolumn]{jpsj2}

\def\runtitle{
  Lattice Distortion and Octupole Ordering Model in Ce$_x$La$_{1-x}$B$_6$}

\def\runauthor{Katsunori \textsc{Kubo} and Yoshio \textsc{Kuramoto}}

\title{
  Lattice Distortion and Octupole Ordering Model in Ce$_x$La$_{1-x}$B$_6$}

\author{Katsunori \textsc{Kubo}\thanks{E-mail: katukubo@cmpt.phys.tohoku.ac.jp}
  and Yoshio \textsc{Kuramoto}}

\inst{Department of Physics, Tohoku University, Sendai 980-8578}

\recdate{April 21, 2003}

\abst
{Possible order parameters of the phase IV in Ce$_x$La$_{1-x}$B$_6$
  are discussed with special attention to the lattice distortion 
  recently observed. 
  A $\Gamma_{5u}$-type octupole order with finite wave number
  is proposed as the origin of the distortion along the [111]  direction.
  The $\Gamma_8$ crystalline electric field (CEF) level splits into
  three levels by 
  a mean field
  with the $\Gamma_{5u}$ symmetry.
  The ground and highest singlets have the same quadrupole moment,
  while the intermediate doublet has an opposite sign. 
  It is shown that any collinear order of $\Gamma_{5u}$-type octupole moment 
  accompanies the $\Gamma_{5g}$-type {\it ferro-quadrupole} order,
  and the coupling of  the quadrupole  moment
  with the lattice induces the distortion.
  The cusp in the magnetization at the phase transition is reproduced,
  but the internal magnetic field due to the octupole moment
  is smaller than the observed one by an order of magnitude.  
}

\kword
{rare-earth hexaboride, Ce$_x$La$_{1-x}$B$_6$, NdB$_6$, NpO$_2$,
  octupole ordering,
  thermal expansion, internal field}

\begin{document}
\sloppy
\maketitle

The multipolar orderings in cubic rare-earth hexaborides
have been studied extensively.
Especially,
CeB$_6$ is a typical material
which shows multipolar orderings.
In CeB$_6$, there are three phases:
the paramagnetic phase (called the phase I),
the antiferro-quadrupole phase (II)
and the antiferromagnetic phase (III).
In Ce$_x$La$_{1-x}$B$_6$, another phase,
so called phase IV, was found 
at $x \simeq 0.75$.~\cite{Sakakibara2}
This phase has attracted much attention,
but the order parameter has not yet been established. 
In the phase IV, 
a large softening of the elastic constant $C_{44}$
was observed.\cite{Suzuki} 
Furthermore the magnetic susceptibility shows a cusp 
at the transition temperature $T_{\text{I-IV}}$ from the phase I,
and the magnetization is almost isotropic at ambient pressure.~\cite{Tayama}

Recently, a small lattice distortion along
the [111] direction was observed in the phase IV.~\cite{Akatsu,Akatsu2}
It is probable that
the softening of $C_{44}$ is associated with this distortion.
It may be tempting to ascribe 
the distortion to the ferro $\Gamma_{5g}$-type quadrupole
order. 
However, the quadrupole moment is not necessarily the primary order parameter.
Large change of the internal field at $T_{\text{I-IV}}$, as probed by
NMR~\cite{Magishi} and $\mu$SR~\cite{Takagiwa}, 
suggests strongly 
that the time reversal symmetry is broken in the phase IV,
which is incompatible with a pure quadrupole order.
Besides, neutron diffraction experiment
found no magnetic reflection in the phase IV.~\cite{Iwasa}
Thus dipole moments are unlikely to be the primary order parameter either,
although the time reversal symmetry is broken.
Therefore, octupole moments,
which break the time reversal symmetry,
become a candidate for the order parameter
in phase IV.~\cite{Kuramoto, Kusunose}
Since the cubic symmetry is broken, and since the anisotropy develops 
in the magnetization under uniaxial pressure,~\cite{Sakakibara,Sakakibara3}
the order parameter should have an anisotropic nature.
Thus the $\Gamma_{5u}$-type, among all octupole moments,
is the most plausible candidate for the order parameter in the phase IV.

Kusunose and Kuramoto~\cite{Kusunose} have already pointed out
using the Ginzburg-Landau (GL) theory that
the $\Gamma_{5u}$ octupole order
with finite wave number
should accompany a ferro-quadrupole moment,
and have suggested a possible lattice distortion. 
However, evaluation of the magnitude of the distortion
is beyond their GL theory.
In this paper, we explore in much greater detail
the consequence of the $\Gamma_{5u}$ octupole order
by the mean field theory,
and propose that the lattice distortion
is due to the order of the $\Gamma_{5u}$ octupole moment.

The CEF ground state of
Ce$^{3+}$ ($J=5/2$) in
Ce$_x$La$_{1-x}$B$_6$ is
the $\Gamma_8$ quartet.~\cite{Zirngiebl,Luthi,Sato}
The excited level $\Gamma_7$ lies about 500K
from the $\Gamma_8$ level
and is neglected.
The $\Gamma_8$ states are represented in terms of eigenstates of $J_z$ as
\begin{align}
  |+ \uparrow \rangle
  &= \sqrt{5/6} |+5/2 \rangle+\sqrt{1/6} |-3/2 \rangle,\label{eq:base1}\\
  |- \uparrow \rangle 
  &= |+1/2 \rangle,\label{eq:base2}
\end{align}
where + and $-$ denote orbital indices,  
and their Kramers partners
$|+ \downarrow \rangle, |- \downarrow \rangle$
are obtained by reversing the signs of $J_z$
in eqs.(\ref{eq:base1}) and (\ref{eq:base2}), respectively.
Within the $\Gamma_8$ quartet,
the number of independent multipolar moments is 15,
and active octupole moments have either of $\Gamma_{2u}$, $\Gamma_{4u}$
or $\Gamma_{5u}$-type symmetry.~\cite{Shiina}
The $\Gamma_{4u}$-type octupole moments
accompany dipole moments,~\cite{Shiina}
and are unlikely to be the order parameter in the phase IV.
The $\Gamma_{2u}$-type octupole moment,
as proposed by ref.\citen{Kuramoto},
does not accompany quadrupole moments.
Therefore it seems difficult to explain
the lattice distortion in the phase IV
by the $\Gamma_{2u}$ octupole orderings.

\begin{figure}[t]
  \includegraphics[width=8cm]{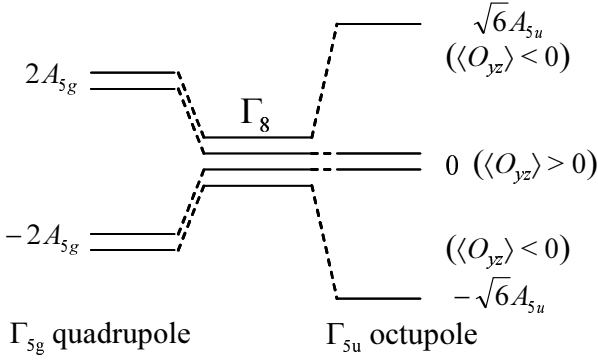}
  \caption{The level scheme in the $\Gamma_{5u}$
  interaction (right) and in the $\Gamma_{5g}$ interaction (left).}
  \label{figure:level}
\end{figure}

Let us introduce pseudospins
$\mib{\sigma}$ and $\mib{\tau}$ to describe the $\Gamma_8$ quartet:
\begin{equation}
  \tau^z|\pm \sigma^z \rangle = \pm|\pm \sigma^z \rangle,
\end{equation}
\begin{equation}
  \sigma^z|\tau^z \uparrow \rangle = +|\tau^z \uparrow \rangle, \ 
  \sigma^z|\tau^z \downarrow \rangle = -|\tau^z \downarrow \rangle,
\end{equation}
and the transverse components which flip the pseudospins.
The $\Gamma_{5u}$ octupole,
$\Gamma_{3g}$ and $\Gamma_{5g}$ quadrupole moments are given
with the notation $(\alpha,\beta,\gamma)=(x,y,z)$, $(y,z,x)$ or $(z,x,y)$
by
\begin{align}
  T^{5u}_\alpha &=
  (\overline{J_\alpha J^2_\beta}-\overline{J^2_\gamma J_\alpha})/(2\sqrt{3})
  =(\zeta^+ \sigma^x,\zeta^- \sigma^y,\tau^x \sigma^z),\\
  O^0_2 &=(2J^2_z-J^2_x-J^2_y)/\sqrt{3} =(8/\sqrt{3})\tau^z,\label{eq:O20}\\
  O^2_2 &=J^2_x-J^2_y=(8/\sqrt{3})\tau^x,\\
  O_{\beta \gamma} &=2 \overline{J_\beta J_\gamma}
  =(2/\sqrt{3})\tau^y \sigma^\alpha,\label{eq:Obc}
\end{align}
where bars on the products represent symmetrization,
\textit{e.g.}, $\overline{J_xJ^2_y} =(J_xJ^2_y+J_yJ_xJ_y+J^2_yJ_x)/3$, 
and we have introduced the notation
$\zeta^{\pm}=-(\tau^x \pm \sqrt{3} \tau^z)/2.$
We find that the `easy axis' of $\Gamma_{5u}$ octupole moment
is along the [111] direction
and equivalent ones, i.e.,
$(|\langle T^{5u}_x \rangle |,
  |\langle T^{5u}_y \rangle |,
  |\langle T^{5u}_z \rangle |) \parallel (1,1,1)$
provided the intersite interaction is isotropic.
Then we consider the following single-site Hamiltonian:
\begin{equation}
  \mathcal{H}_{\text{single-site}}=A_{5u}(T^{5u}_x+T^{5u}_y+T^{5u}_z),
  \label{eq:local-model}
\end{equation}
where $A_{5u}$ denotes the octupolar mean field.
The energy levels
of this Hamiltonian are shown in the right part of 
Fig.~\ref{figure:level}.
The $\Gamma_8$ level splits into
three levels.
Not only the time reversal symmetry,
but also the cubic symmetry are broken.
The multipolar moments in
the ground and highest
states are
$\langle T^{5u}_x \rangle=\langle T^{5u}_y \rangle
=\langle T^{5u}_z \rangle=\mp \sqrt{2/3}$,
$\langle O_{yz} \rangle=\langle O_{zx} \rangle
=\langle O_{xy} \rangle=-2/3$
and the others are zero.
Thus,
ferro, antiferro and other collinearly
ordered states with
$(\langle T^{5u}_x \rangle,
  \langle T^{5u}_y \rangle,
  \langle T^{5u}_z \rangle ) \parallel (1,1,1)$
have a homogeneous $\Gamma_{5g}$ moment
and the crystal distorts along  [111].

Obviously a $\Gamma_{5g}$-type ferro-quadrupole interaction
can also lead to
$\langle O_{yz} \rangle =\langle O_{zx} \rangle =\langle O_{xy} \rangle \ne 0$.
In this case the energy level splits into two levels,
and the ground state is two-fold degenerate as shown in the left part of 
Fig.~\ref{figure:level}.
Experimentally,
the entropy changes much at the transition from the phase I to IV,
and little
from the phase IV to III in Ce$_{0.75}$La$_{0.25}$B$_{6}$.~\cite{Furuno,Suzuki}
Thus it is likely that the degeneracy of
the each $f$-electron state is already lifted in the phase IV.
This situation also makes it hard for
the $\Gamma_{5g}$ quadrupoles to be the order parameter in the phase IV.

From the above discussion,
we find that 
the order parameter which
breaks the time reversal symmetry 
and induces quadrupole moments,
but which accompanies no dipole moment
is only the $\Gamma_{5u}$ octupole moment.
However, we cannot determine
the periodicity of the $\Gamma_{5u}$
ordered state in the phase IV
only from the above consideration,
because any collinearly ordered state with
$(\langle T^{5u}_x \rangle,
  \langle T^{5u}_y \rangle,
  \langle T^{5u}_z \rangle ) \parallel (1,1,1)$
has a uniform $\Gamma_{5g}$ moment.
To carry out the mean field theory explicitly,
we consider a G-type antiferro-octupole order
as the simplest example of $\Gamma_{5u}$ orders.
The importance of $\Gamma_{5u}$ nearest-neighbor interaction
in causing the change from the phase III to III$^{\prime}$,
even with weak magnetic field, 
was pointed out in ref.\citen{Kusunose},
and we consider only this nearest-neighbor interaction.
We take the following model:
\begin{equation}
  \mathcal{H}=I^{5u} \sum_{(i,j)}
  \sum_{\alpha=x,y,z}T^{5u}_{\alpha \ i} T^{5u}_{\alpha \ j},
\end{equation}
where $(i,j)$ denotes a nearest-neighbor pair.
We study this Hamiltonian by the mean field theory, and
choose the value of $I^{5u}$ so as
to reproduce the transition temperature
$T_{\text{I-IV}}$ in Ce$_{0.75}$La$_{0.25}$B$_6$,
i.e., $T_{5u}=6I^{5u}=1.7$K.

\begin{figure}[t]
  \includegraphics[width=8cm]{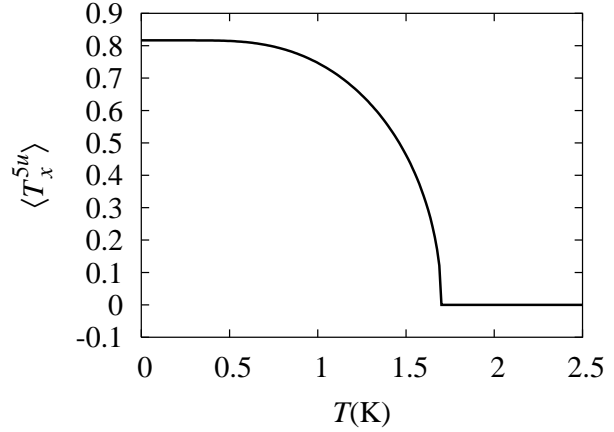}
  \caption{Temperature dependence of
    the antiferro-octupole moment
    $\langle T^{5u}_x \rangle$
    $(=\langle T^{5u}_y \rangle =\langle T^{5u}_z \rangle )$.}
  \label{figure:Tbx}
\end{figure}

\begin{figure}[t]
  \includegraphics[width=8cm]{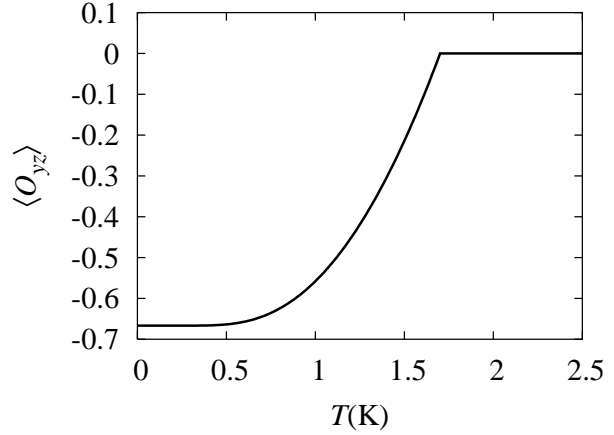}
  \caption{Temperature dependence of
    the ferro-quadrupole moment
    $\langle O_{yz} \rangle$
    $(=\langle O_{zx} \rangle =\langle O_{xy} \rangle )$.}
  \label{figure:Oyz}
\end{figure}

The temperature dependence of
$\langle T^{5u}_x \rangle$ and $\langle O_{yz} \rangle$
are shown in Figs.~\ref{figure:Tbx} and ~\ref{figure:Oyz}.
The solution obtained has the antiferro-octupole order with
$\langle T^{5u}_x \rangle=\langle T^{5u}_y \rangle =\langle T^{5u}_z \rangle$,
accompanying ferro-quadrupole moment
$\langle O_{yz} \rangle =\langle O_{zx} \rangle =\langle O_{xy} \rangle$.
We also find other equivalent solutions:
$(\langle T^{5u}_\alpha \rangle,
\langle T^{5u}_\beta \rangle,
\langle T^{5u}_\gamma\rangle,
\langle O_{\beta \gamma}\rangle,
\langle O_{\gamma \alpha}\rangle,
\langle O_{\alpha \beta}\rangle)
= (\pm B,\pm B,\mp B,+C,+C,-C)$
where quantities $B$ and $C$ depend on temperature.

\begin{figure}[t]
  \includegraphics[width=8cm]{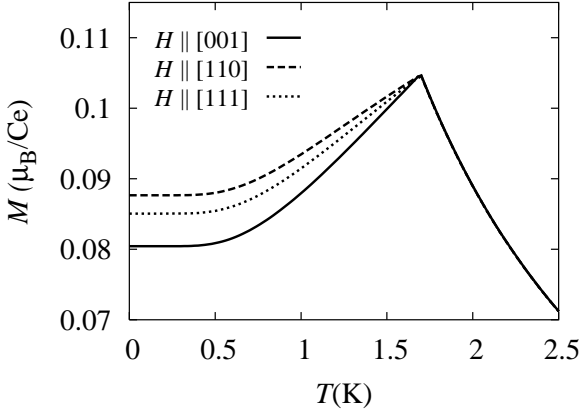}
  \caption{Temperature dependence of the magnetization
    in magnetic field $H=0.2$T along various directions.}
  \label{figure:M}
\end{figure}

In Fig.~\ref{figure:M},
we show the magnetization
in magnetic field $H=0.2$T
along three high-symmetry directions.
The magnetization has a cusp at $T_{5u}$,
which is consistent with experimental observation.~\cite{Tayama}
We have also examined the case where
the pure $\Gamma_{5g}$ quadrupole moment is
the order parameter, and found that 
the magnetization changes little
at the quadrupole transition temperature.
The magnetization below $T_{5u}$ is anisotropic.
One should not, however, take this anisotropy seriously since other interactions such as quadrupole and dipole interactions also influence the anisotropy.
In fact 
the easy axis in the mean field theory
is different from that observed
in the phase IV under uniaxial pressure and
in the phase III.

We now consider lattice distortion
in the antiferro-octupole ordered state.
The elastic energy associated with the $\Gamma_{5g}$ moments is given 
per unit volume by
\begin{equation}
  E = \sum_{\alpha \beta=yz,zx,xy}
  \left(
    2\epsilon^2_{\alpha \beta}C^{(0)}_{44}
    +g_{\Gamma_5} \sum_{i}
    \epsilon_{\alpha \beta} \langle O_{\alpha \beta \ i} \rangle
  \right),
\end{equation}
where 
$\epsilon_{\alpha \beta}$ is the strain tensor,
$g_{\Gamma_5}$ is the magneto-elastic coupling constant, and
$C^{(0)}_{44}$ is the (bare) elastic constant.
The sum runs over Ce sites $i$ in the unit volume.
By minimizing the elastic energy, we obtain
\begin{equation}
  \epsilon_{\alpha \beta} =-\frac{g_{\Gamma_5}}{4 C^{(0)}_{44}}
  \sum_{i} \langle O_{\alpha \beta \ i} \rangle.
\end{equation}
We use the following experimental values for Ce$_{0.75}$La$_{0.25}$B$_6$: 
$|g_{\Gamma_5}|=155$K,
$C^{(0)}_{44} \simeq 8.2 \times 10^{11} \text{erg/cm}^3$,~\cite{Suzuki}
and the lattice constant $a=4.13$\AA.
At absolute zero, the magnitude of quadrupole moments is given by
$|\langle O_{yz \ i} \rangle|=|\langle O_{zx \ i} \rangle|
=|\langle O_{xy \ i} \rangle|=2/3$,
and we obtain
\begin{equation}
  |\epsilon_{yz}|=|\epsilon_{zx}|=|\epsilon_{xy}|=4.6 \times 10^{-5}.
  \label{eq:strain}
\end{equation}

In order to account for the observed lattice contraction along [111],
we consider the two possibilities: (i) $g_{\Gamma_5}<0$,
and (ii) $g_{\Gamma_5}>0$. 
In the case (i), 
a slight stress accompanying the measurement
breaks the equivalence of four octupole domains, and choose a domain
for which the contraction becomes the maximum along [111].  Namely we have 
\begin{equation}
\Delta l/l=2\epsilon_{yz}=-9.3 \times 10^{-5},
  \label{eq:distortion2}
\end{equation}
with $\langle O_{yz \ i} \rangle=\langle O_{zx \ i} \rangle
=\langle O_{xy \ i} \rangle= -2/3$.
Note that along directions $[\bar{1}11] ,[1\bar{1}1]$,
and $[11\bar{1}]$ the lattice should expand by 
\begin{equation}
  \Delta l/l=2|\epsilon_{yz}|/3=3.1 \times 10^{-5}.
  \label{eq:distortion}
\end{equation}
On the other hand, in the case (ii),
a positive stress along [111] may favor a domain which contracts along this direction. 
Then the contraction is given by eq.(\ref{eq:distortion}) with the minus sign.
In the single domain with
$-\langle O_{yz} \rangle =\langle O_{zx} \rangle =\langle O_{xy} \rangle$
for example, 
an expansion along $[\bar{1}11]$ should be present,
and its magnitude is three times larger than the contraction along [111].

A magnetic field induces
antiferromagnetic moments which tend to be perpendicular to $\mib{H}$.
Hence with $\mib{H} \parallel$ [111] without external stress, 
a state where one of $\langle O_{\alpha\beta} \rangle$'s
has a sign different from the others is stabilized.
Experimentally the contraction is enhanced under $\mib{H} \parallel$ [111],
which alone favors the case (ii) in our model.
In the actual system, diagonal components
$\epsilon_{\alpha\alpha}$ becomes positive in the phase IV.
This comes from the volume strain which is not included in our model.
According to experiment~\cite{Akatsu2}, the shear strain
with the assumption of the trigonal symmetry around [111],
i.e., the case (i),
is derived as
$\epsilon_{\alpha\beta}=-4 \times 10^{-6}$ without magnetic field
at 1.3K.
Assuming that the phase IV is stable down to zero temperature, 
$\epsilon_{\alpha\beta}$ extrapolate to
$(-6\sim -10)\times 10^{-6}$ at absolute zero.
The absolute value is by an order of magnitude smaller than
our estimate in eq.(\ref{eq:strain}).
Then the observed reduction of 
$|\langle O_{\alpha\beta} \rangle|$ 
should come from quantum fluctuations neglected here.
Another possibility is that the case (ii) is realized.
In this case three times larger $|\epsilon_{\alpha \beta}|$, i.e.,
$(2\sim 3)\times 10^{-5}$,
is obtained from the same experimental result~\cite{Akatsu2}.
Our model 
in the case (ii) predicts
a larger lattice contraction 
along the [110] direction with $\mib{H} \parallel [110]$
than that along the [111] direction
with $\mib{H} \parallel [111]$.

For comparison,
we estimate in the same way the magnitude of the lattice distortion
of NdB$_6$ in the $O^0_2$ ferro-quadrupole ordered state
by using the experimental values:
$|g_{\Gamma_3}|=220$K,
$(C_{11}-C_{12})/2
\simeq 20 \times 10^{11} \text{erg/cm}^3$,~\cite{Tamaki}
$a=4.12$\AA, \
and the Lea-Leask-Wolf parameter $x_{\text{LLW}}=-0.82$.~\cite{Pofahl}
We assume that $(C^{(0)}_{11}-C^{(0)}_{12})/2$
is almost the same as the observed
$(C_{11}-C_{12})/2$.
In a magnetic field $\mib{H} \parallel$ [100],
a state with the antiferromagnetic moment perpendicular to [100]
is stabilized.
For a domain where the magnetic moment is along [001],
we obtain $|\Delta l|/l=2.8 \times 10^{-4}$ along $\mib{H}$. 
This value compares favorably with the experimental one
$\Delta l/l \simeq -2 \times 10^{-4}$ at 2K and $H=2.1$T.\cite{Sera}
The distortion in NdB$_6$ is by an order of magnitude larger 
than that in eqs.(\ref{eq:distortion2}) and (\ref{eq:distortion}).
The reason is that the quadrupole moment
$|\langle O^0_2 \rangle| =4.5$ in NdB$_6$
is much larger than the corresponding value
$|\langle O_{yz} \rangle|=2/3$
in the octupolar state of Ce$_x$La$_{1-x}$B$_6$,
while the magneto-elastic coupling constants are of the same order.

We now discuss
the internal magnetic field associated with the octupole order.
The $\mu$SR time spectra in the phase IV
consist of a Gaussian component
and an exponential component.~\cite{Takagiwa}
The observation of a Gaussian relaxation
indicates that
internal fields are randomly distributed,
and/or fluctuating.
The internal field deduced from the Gaussian relaxation
is the order of 0.1T.
We discuss whether
the octupole moment can be the origin
of the Gaussian relaxation.
In $\mu$SR measurement,
$\mu^+$ locates at $(a/2,0,0)$ and equivalent sites~\cite{Amato,Kadono,Schenck}
with a Ce ion chosen as the origin.
As a reference the internal field from
a Bohr magneton $\mu_B$ is estimated to be
$H_{\text{dipole}} = \mu_B/(a/2)^3 \sim 0.1 \text{T},$
with $a/2\sim 2$\AA.
The internal field from an octupole with the size $r$ is estimated to be
$H_{\text{octupole}} = \mu_B r^2/(a/2)^5 \sim 0.01 \text{T},$
with $r \sim a_B=0.53$\AA.
For a more accurate estimate, we consider the multipole expansion
of the vector potential from local electrons as given by~\cite{Schwartz}
\begin{equation}
  \mib{A}(\mib{r})=\sum_{k,m}
  \frac{-\text{i}}{k} r^{-(k+1)}
  \left(\mib{l}C^{(k)}_{m}(\theta,\phi)\right)M^{m}_{k},
\end{equation}
where
$\mib{l}$ is the orbital angular momentum operator,
$C^{(k)}_{m}(\theta,\phi)$
is $\sqrt{4\pi/(2k+1)}$ times
the spherical harmonics $Y_{k m}(\theta,\phi)$,
and $M^{m}_{k}$ is the magnetic multipole moment.
The multipole moment
is determined by 
the wave function $\psi_i(\mib{r})$ of the $i$-th electron,
the orbital and spin angular momentum operators
$\mib{l}_i$ and $\mib{s}_i$.
Namely we have
\begin{equation}
  \begin{split}
    M^{m}_{k}=&\mu_{\text{B}}
    \sum_i \int \text{d}\mib{r}_i
    \psi^*_i(\mib{r}_i)
    \left(\mib{\nabla}_i r^k_i C^{(k) *}_{m}(\theta_i,\phi_i)\right)\\
    &\cdot 
    \left(\frac{2}{k+1}\mib{l}_i+2\mib{s}_i\right)
    \psi_i(\mib{r}_i).
  \end{split}
  \label{eq:Mmu}
\end{equation}
Eq.(\ref{eq:Mmu}) is evaluated
with use of the operator equivalents method.~\cite{Inui}
For our purpose it is sufficient to
consider only octupole moments.
Then we obtain $M^{m}_3$ through calculation of the reduced matrix element
of the third rank tensor.
The result for one electron states with $J=5/2$, $L=3$, $S=1/2$ is given by
\begin{equation}
  M^{m}_3=-\frac{2}{35}\mu_{\text{B}}
  \langle r^2 \rangle
  \langle J^{(3)}_{m} \rangle,
\end{equation}
where
the third-rank tensor operators
$J^{(3)}_{m}$ are defined in ref.~\citen{Shiina}.

Freeman and Desclaux obtained the estimate
$\langle r^2 \rangle =1.298$ in atomic unit
by a relativistic Dirac-Fock calculation.~\cite{Freeman}
By using this value, we obtain about 40G
as the magnitude of internal field
at $(a/2,0,0)$ from a Ce ion.
This value is by an order of magnitude smaller than that derived
by the $\mu$SR measurement.~\cite{Takagiwa}
Thus the static octupolar moment alone cannot account for
the Gaussian relaxation of the $\mu$SR spectra.
In a future work, we plan to study in more detail fluctuations
in the octupole ordered state.

We mention that Paix\~{a}o \textit{et al.} have  proposed
for the ordered state of NpO$_2$
that a triple-$\mib{q}$ $\Gamma_{5u}$-type octupole ordering is realized
and a triple-$\mib{q}$ $\Gamma_{5g}$ quadrupole
moment is induced.~\cite{Paixao}
We estimate
the internal magnetic field in NpO$_2$ to be about 80G
at 8K from
the observed 
muon spin precessing frequency of 7MHz.~\cite{Kopmann}
The internal field is of the same order
as our estimate for Ce$_x$La$_{1-x}$B$_6$.

In the antiferro-octupolar phase, antiferromagnetism should be induced
by uniaxial stress.~\cite{Kuramoto}
In the following we estimate the magnitude of the induced moment.
The uniaxial pressure $p$ along the [001] direction accompanies 
the $\Gamma_3$ strain 
\begin{equation}
  (C_{11}-C_{12}) (2\epsilon_{zz}-\epsilon_{xx}-\epsilon_{yy})=2p.
\end{equation}
We have solved the mean field equation with the finite
$\Gamma_3$ strain 
and the corresponding quadrupole-strain interaction $g_{\Gamma_3}$.
The antiferromagnetic moment is induced in the $xy$-plane
since $\langle T^{5u}_x \rangle \neq 0$
together with $\langle J^2_y-J^2_z \rangle\neq 0$ gives
$\langle J_x \rangle\neq 0$.
The direction of the antiferromagnetic moment
is along the [110] or $[1\bar{1}0]$.
With experimental values:
$(C_{11}-C_{12})/2 \simeq 20.4 \times 10^{11}$(erg/cm$^3$)
and $|g_{\Gamma_3}|=120$K,~\cite{Suzuki}
the mean-field solution for $\langle T^{5u}_z \rangle$ becomes zero
for $p \geq 0.9$GPa.
This implies the octupole and dipole moments both lying in the $xy$-plane.
With $p=1$GPa, the magnetic moment
is estimated to be 0.88$\mu_{\text{B}}$ (0.82$\mu_{\text{B}}$),
if $g_{\Gamma_3}$ is positive (negative).
This value should actually be reduced
by quantum fluctuations.
However, it is likely that the magnitude remains within experimental access.

To summarize,
we have proposed that
the $\Gamma_{5u}$ octupole moment
is a plausible candidate for the order parameter of the phase IV.
This ordered state is consistent with
the lattice distortion along the [111] direction,~\cite{Akatsu,Akatsu2}
the broken time reversal symmetry,~\cite{Magishi,Takagiwa}
no dipole moment found~\cite{Iwasa},
and the cusp in the magnetization at $T_{\text{I-IV}}$.~\cite{Tayama}
However, the internal field estimated by our mean field theory
is much smaller than that suggested by
the $\mu$SR experiment.~\cite{Takagiwa}
We recall that the NMR spectra become broad in the phase IV,~\cite{Magishi}
and the spectra cannot be explained
by the static and staggered octupole moments either.
Therefore clarifying dynamical aspects and identifying the  periodicity in
the phase IV are challenging open problems to be addressed in the near future.

This work has been supported partly by
Special Coordination Funds for Promoting Science and
Technology, and by the NEDO international collaboration program
``New boride materials''.

\end{document}